\documentstyle[12pt]{article}
 
\def\be{\begin{equation}}
\def\ee{\end{equation}}
\def\a{\alpha}
\def\s{\sigma}
\def\c{\tilde{c}}
\def\b{\tilde{b}}
\def\n{\{n\}}
\def\nn{\hat{n}} 
\def\O{\hat{O}}
\def\Ot{\hat{\tilde{O}}} 
\def\F{\hat{F}}
\def\l{\lambda} 
\def\x{x}
\def\vac{|0\rangle} 
\def\aa{{\large a}}
\def\ra{\rangle}
\def\la{\langle}
\def\ln{\mbox{ln}}

\begin{document}

\begin{center} 
{\large Construction of Monodromy Matrix in the F - basis and scalar \\
        products in Spin chains. }
\end{center}

\vspace{0.1in} 

\begin{center}
{\large  A.A. Ovchinnikov  }
\footnote{E-mail address: ovch@ms2.inr.ac.ru}
\end{center}

\begin{center}
{\it Institute for Nuclear Research, RAS, Moscow.}
\end{center}

\vspace{0.1in}

\begin{abstract}
We present in a simple terms the theory of the factorizing operator 
introduced recently by Maillet and Sanches de Santos for the spin- 1/2 
chains. We obtain the explicit expressions for the matrix elements 
of the factorizing operator in terms of the elements of the 
Monodromy matrix. We use this results to derive the expression 
for the general scalar product for the quantum spin chain.  
We comment on the previous determination of the scalar product of 
Bethe eigenstate with an arbitrary dual state. We also establish 
the direct correspondence between the calculations of scalar 
products in the F - basis and the usual basis.

\end{abstract}

\vspace{0.3in}

              {\bf 1. Introduction.}

\vspace{0.2in}

One of the most important open problems in the theory of quantum 
integrable models is the calculation of the correlation functions. 
In the framework of the Algebraic Bethe Ansatz method \cite{FST} 
the problem is the combinatorial complexity of calculations due to 
the structure of Bethe eigenstates. 
Only the very limited number of
physical results for the correlation functions 
(or formfactors) of integrable models 
have been obtained from the first principles.

  The concept of factorizing F - matrix was recently introduced by 
Maillet and Sanches de Santos \cite{MS} following the concept of Drinfeld's 
twists in his theory of Quantum Groups. 
The existence of the factorizing matrix allows one by means of similarity 
transformation to make the elements of the Monodromy matrix totally symmetric 
with respect to an arbitrary  permutation of indices $1,\ldots N$ together 
with the corresponding inhomogeneity parameters $\xi_{1},\ldots \xi_{N}$. 
The matrix $F_{1...N}$ is defined as follows. 
Consider any permutation $\sigma\in S_N$.  
Then any component in the auxiliary space $0$ of the monodromy matrix $T_0(t)$
depending on the parameter $t$  transforms as 
\[
T_{0,\s 1\ldots \s N}=R^{\s}_{1\ldots N} T_{0,1\ldots N} 
(R^{\s}_{1\ldots N})^{-1}, 
\]
where $R^{\s}_{1...N}$ is the operator constructed from the elementary $S$- 
matrices.  Then if the matrix $F_{1...N}$ is defined according to 
\be 
(R^{\s}_{1\ldots N})^{-1}= (F_{1\ldots N})^{-1} F_{\s 1\ldots \s N}, 
\label{fff}
\ee
in the F- basis the monodromy matrix $T_{0}^{F}=F T_{0} F^{-1}$ is totally 
symmetric: 
\[
T_{0,\s 1...\s N}^{F}(t) = T_{0,1...N}^{F}(t), ~~~~~\s \in S_N  
\] 
for arbitrary $t$. 
It was also realized that there exist the factorizing matrix F which 
diagonalize the $A$ ($D$)- component of the monodromy matrix. In this basis 
the operators $B$ and $C$ have quite a simple quasilocal form which recently   
made possible the direct computations for the important class of the (general)
correlation functions \cite{KMT}. 
In some respect this approach allows one to simplify the calculations 
in comparison with the general theory of scalar products \cite{K1}, \cite{IK} 
and gives an alternative way to obtain many important results for the XXZ spin 
chain. Among them one 
can mention the derivation by Korepin the Gaudin formula for the norm of Bethe 
eigenstates \cite{K1} and the calculation of the scalar product of Bethe 
eigenstate with an arbitrary dual state \cite{S}, which leads straightforwardly 
to the determinant representation of the formfactors of basic (local) operators  
(see e.g. \cite{Kojima}, \cite{KS} for the analogous calculations 
for the case of Bose-gas with $\delta$ - function interaction).     
Using the Algegraic Bethe Ansatz and the solution of the quantum inverse
scattering problem the authors of ref.\cite{KMT1} obtained the multiple
integral representation for the correlation functions found previously 
using the other methods (see for example \cite{M} and references therein). 

At the same time in the approach of ref.\cite{MS} the construction of 
the F - matrix itself and of the matrix elements of different operators 
in the F - basis is quite complicated. The construction involves the 
specially defined partial F - matrices and some steps (for example the proof 
that the factorizing matrix diagonalize the $A$ ($D$)- operators) are based on a 
special recurrence procedures. Also the explicit expressions for the matrix 
elements of the operator F, which can be used in practical computations,  
was not presented. The generalization of the 
construction to the case of the other models is not straightforward. 
  
The aim of the present paper is twofold. 
First, we present in a simple 
terms the theory of the factorizing operator introduced in ref.\cite{MS} 
and derive the new formulas for the matrix elements of the factorizing 
operator (Section 2). 
In comparison to the authors \cite{MS} we proceed in a different way. 
First of all we construct the matrix that diagonalizes the operator $A(t)$. 
We obtain the simple expression for its matrix elements and  the matrix elements 
of the inverse operator in terms of the matrix elements of the products of
the operators $B(t)$, $C(t)$. These expressions have not been obtained previously.  
Then we find the expressions for the other 
operators $B$, $C$ in the new basis (Section 3). 
Afterwards we prove that the constructed operator 
is the factorizing operator as defined above. 
This procedure allows one to obtain the simple expressions for the matrix elements
of the factorizing operator which can be used in practical computations. 
The presented formalism could be usefull for the other integrable models 
(for example the generalization to the case of XYZ- spin chain is straightforward).   

Second, we apply the developed formalism to the  calculation of the general 
scalar product in a way which is different from that of ref.\cite{KMT} (Section 4).   
We obtain the general expression for the scalar product of ref.\cite{K1} in 
a direct (i.e. without any recurrence procedures) and a simple way. In 
some sense our calculation clarifies the mathematical structure underlying 
the derivation of ref's \cite{K1}, \cite{IK}. The key point in our derivation 
is the formulas for the matrix elements of the factorizing operator obtained in 
Section 2. 
Next, in Section 4, we supplement the derivation \cite{S} of the determinant
representation of the scalar product of Bethe eigenstate with an arbitrary dual 
state. We present here the new version of the proof given by Slavnov \cite{S}, 
which is based on the recurrence relations. 
We also establish the direct correspondence between the calculations of 
an arbitrary scalar products in the F - basis and the conventional basis 
in the Appendix C. The calculation of the sums for the scalar products 
presented here can serve as an independent proof of the formulas 
for the matrix elements of the Monodromy matrix in the F - basis.

\vspace{0.2in}

     {\bf 2. Construction of the factorizing operator.}  

\vspace{0.2in}

    We consider in this paper the XXX or XXZ spin- 1/2 chains of finite 
length N. Before dioganalizing the operator $A$,  
let us fix the notations: the normalization of basic S - matrix, 
the definition of monodromy matrix and write down the Bethe Ansatz equations. 
For the rational case (XXX- chain) the S- matrix has the form 
$S_{12}(t_1,t_2)=t_1-t_2+\eta P_{12}$, where $P_{12}$ is the permutation 
operator. In general (XXZ) case it can be written as 
\[
S_{12}(t_1,t_2) =  \left( 
\begin{array}{cccc}  
a(t) & 0 & 0 & 0 \\
0 & c(t) & b(t) & 0 \\
0 & b(t) & c(t) & 0 \\
0 & 0 & 0 & a(t) 
\end{array}  \right)_{(12)}, ~~~~ t=t_1-t_2. 
\]
One can choose the normalization $a(t)=1$ so that the functions $b(t)$ and $c(t)$ 
become:  
\[
\b(t)=\phi(\eta)/\phi(t+\eta)~~~~~~~\c(t)=\phi(t)/\phi(t+\eta), 
\]
where $\phi(t)=t$ for the isotropic (XXX) chain and $\phi(t)=\sinh(t)$ 
for the XXZ- chain. With this normalization the $S$-matrix satisfies 
the unitarity condition $S_{12}(t_1,t_2)S_{21}(t_2,t_1)=1$. 
The monodromy matrix is defined as 
\[
T_{0}(t,\{\xi\})= S_{10}(\xi_1,t)S_{20}(\xi_2,t)...S_{N0}(\xi_N,t), 
\]
where $\xi_i$ are the inhomogeneity parameters. We define the operator entries 
in the auxiliary space $(0)$ as follows: 
\[
\la \beta |T_0|\a \ra = \left( 
\begin{array}{cc}
A(t) & B(t) \\
C(t) & D(t) 
\end{array} \right)_{\a\beta}; ~~~~\a,\beta = (1,2) = (\uparrow; \downarrow). 
\]
We denote throughout the paper $(\uparrow;\downarrow)=(1;0)$ so that the 
pseudovacuum (quantum reference state) $\vac=|\{00...0\}_N\ra$. 
The triangle relation (Yang-Baxter equation) reads:  
\[
S_{12}S_{13}S_{23}=S_{23}S_{13}S_{12},~~~R_{00^{\prime}}T_0 T_{0^{\prime}}= 
T_{0^{\prime}}T_0 R_{00^{\prime}}; ~~~
R_{00^{\prime}}=P_{00^{\prime}}S_{00^{\prime}}. 
\]
The action of the operators on the pseudovacuum is: $A(t)\vac=\aa(t)\vac$ 
($\aa(t)=\prod_{\a}\c(\xi_{\a}-t)$), $D(t)\vac=\vac$, $C(t)\vac=0$. 
The Bethe Ansatz equaions for the eigenstate of the Hamiltonian
$\prod_{i=1}^M B(t_i)\vac$ are 
\[
\aa(t_i)=\prod_{\a\neq i} \c(t_{\a}-t_i) (\c(t_i-t_{\a}))^{-1}, 
\]
and the corresponding eigenvalue of the transfer - matrix $Z(t)=A(t)+D(t)$ is 
\[
\Lambda(t,\{t_{\a}\})=\aa(t)\prod_{\a=1}^{M}\c^{-1}(t_{\a}-t)+ 
\prod_{\a=1}^{M}\c^{-1}(t-t_{\a}),   
\]
where $t_{\a}$ are the solution of Bethe Ansatz equations.

 In order to construct the operator $\O = \O_{1...N}$ which diagonalizes the 
operator $A(t)$ ($\O^{-1} A \O =diag(A)$), let us first construct the 
eigenfunctions of the operator $A$. One can do it in two different ways. 
First, note that $A$ - is a triangular matrix in a sense that it makes particles 
(the spin-up - coordinates) move to the right on the lattice $1...N$. 
Thus its eigenvalues are coincide with its diagonal matrix elements and 
therefore are charactarized by the set of integers ($n_1,...n_M$) - the spin-up 
positions. Let us denote this eigenfunctions by $|\phi (n_1,...n_M)\ra$. Clearly, 
$|\{00..0\}_{N-M}\{11..1\}_{M}\ra$ is an eigenstate of $A(t)$. 
Therefore, considering the permutation 
\[
S_{10}...S_{N0} \rightarrow  
S_{10}...S_{N0}S_{n_M 0}...S_{n_1 0} ~~~~(n_1 < n_2 <...< n_M),
\]
we realize that 
\be
|\phi (n_1,...n_M)\ra = T_{n_1}T_{n_2}...T_{n_M} |n_1,n_2,...n_M\ra 
\label{tn}
\ee
where 
\[
T_n = S_{n+1,n}S_{n+2,n}...S_{Nn}, 
\]
is an eigenstate of the operator $A(t)$. Note that if we modify the given 
permutation interchanging $n_i\leftrightarrow n_j$ the operator in the right-hand 
side of (\ref{tn}) modifies but the state $|\phi(\n)\ra$ remains the same. For example,
for $M=2$: $T_{n_2}T_{n_1}^{\prime}=T_{n_1}T_{n_2}S_{n_{2}n_{1}}^{-1}$, where $n_1<n_2$ 
and the prime means the absence of the term $S_{n_{1}n_{2}}$ in $T_{n_1}$. Since 
$S_{n_{1}n_{2}}|n_1,n_2\ra=|n_1,n_2\ra$ the state remains the same.

The second way - is to consider the state 
\be
|\phi (n_1,...n_M)\ra = B(\xi_{n_1})B(\xi_{n_2})...B(\xi_{n_M}) \vac,~~~(n_i \neq n_j). 
\label{bb}
\ee
Using the fundamental commutation relation:
\be
A(t)B(q)= \frac{1}{\c(q-t)}B(q)A(t) - \frac{\b(q-t)}{\c(q-t)}B(t)A(q),
\label{ab}
\ee
and the fact that for the pseudovacuum state $A(\xi_i)\vac =0$, we find again 
that $|\phi(\n)\ra$ ($\n=\{n_1,...n_M\}$) is an eigenstate of $A(t)$ with the 
eigenvalue $A_{\n\n}(t)=\prod_{\a\neq n_k}\c(\xi_{\a}-t)$, and the operator 
$A(t)$ in the new basis has the following form: 
\be
A^{F}(t)= \prod_{\a=1}^N 
\left( \begin{array}{ll}
\c(\xi_{\a}-t), & \a \neq n_k \\
    1,          & \a = n_k 
\end{array} \right) 
\label{a}
\ee

One can see that the states $|\phi(\{n\})\ra$ (\ref{tn}) and (\ref{bb}) are coincide. 
This can be seen using the identity 
\be
B(\xi_n)=T_n \la 0|S_{10}\ldots S_{n-1,0}P_{n0}|1\ra 
\label{bxi}
\ee
and ordering the (commuting) operators $B(\xi_{n_i})$ in eq.(\ref{bb}) 
according to the prescription $n_1<n_2<...<n_M$, so that the second operator 
in the last formula simply creates the particle (spin- up) at the site $n$ 
with the amplitude equal to unity.   

  Now we can introduce the diagonalizing operator $\O=\O_{1...N}$ (we will 
see later that it is also the factorizing operator). Let us define the 
operator $\O$ such that 
\[
|\phi(\{n\})\ra = \O(\{n\})|\{n\}\ra = \O |\n\ra , 
\]
where $\O(\{n\})$ is given by the equation (\ref{tn}). Clearly the operator 
$A^{F}(t)=\O^{-1}A(t)\O$ - is diagonal (see (\ref{a})). Now we immediately get 
the simple formula for the matrix elements of the factorizing operator: 
\be
\O_{\{m\}\{n\}} = \la\{m\}|B(\xi_{n_1})B(\xi_{n_2})...B(\xi_{n_M}) \vac, 
\label{me}
\ee 
where $\{m\}=\{m_1,...m_M\}$. From this expression (see (\ref{bxi})) one 
see that $\O$ is the triangular matrix (in the same sense as the operator 
$A((t)$). 
Let us show that the operator $\O$ is invertible and construct the inverse 
operator. To find this operator we have to consider the following dual states:
 \be
\la\tilde{\phi} (n_1,...n_M)| = \la 0| C(\xi_{n_1})C(\xi_{n_2})...C(\xi_{n_M}). 
\label{cc}
\ee
Define the operator $\Ot$ analogously to the previous case as 
$\la\tilde{\phi}(\{n\})|=\la\{n\}|\Ot$. The matrix elements of this operator 
are equal to 
\be
\Ot_{\{m\}\{n\}} = \la 0|C(\xi_{m_1})C(\xi_{m_2})...C(\xi_{m_M})|\n\ra. 
\label{mec}
\ee  
Let us calculate the following scalar product  
\be
\la\tilde{\phi}(\{m\})|\phi(\{n\})\ra = \la \{m\}|\Ot\O|\{n\}\ra =
\label{oo}
\ee
\[
\la 0|C(\xi_{m_1})C(\xi_{m_2})...C(\xi_{m_M})
B(\xi_{n_1})B(\xi_{n_2})...B(\xi_{n_M})\vac. 
\]
This scalar product can be calculated using another well known 
relation relation between the elements of the monodromy matrix 
following from the Yang-Baxter equation:   
\be
\left[ B(q),C(t)\right]=\frac{\b(t-q)}{\c(t-q)}\left( D(q)A(t)-D(t)A(q)\right),   
\label{bc}
\ee
and using again that for any site $i$, $A(\xi_i)\vac=0$.
Moving  the operators $A$ and $D$ to the right, 
repeating consequently the relation (\ref{ab}) and the analogous relation for 
$D$ (which differs from eq.(\ref{ab}) by the interchange $t\leftrightarrow q$ 
in the coefficients between the products of the operators)  
and using the equations 
\[
C(\xi_n)B(\xi_n)\vac =\left(\prod_{\a\neq n}\c(\xi_{\a}-\xi_{n})\right)\vac, 
\]
we obtain that the matrix $\hat{f}$ is diagonal : 
\[
\la\tilde{\phi}(\{m\})|\phi(\{n\})\ra = \la\{m\}|\Ot\O|\{n\}\ra =
\la \{m\}|\hat{f}|\{n\}\ra =
\left(\prod_{i}\delta_{n_i,m_i}\right)f(\{n\}),  
\]
The corresponding diagonal matrix elements can be found either using the 
procedure mentioned above or, which is the simplest way, using the 
representation (\ref{bxi}) (and the same for the operator $C(t)$). 
In fact, taking into account the triangularity of $\O$ and $\Ot$ we obtain:  
\[
\la \{n\}|\hat{f}|\{n\}\ra =\sum_{\{m\}}\Ot_{\n\{m\}}
\O_{\{m\}\n}= \Ot_{\n\n}\O_{\n\n}.   
\]
The diagonal matrix elements can be easily calculated 
using the representation (\ref{me}) and the formula (\ref{bxi}). We get: 
\be
f(n_1,...n_M)=\prod_k\left(\prod_{\a\neq n_k,n_j}\c(\xi_{\a}-\xi_{n_k})\right) . 
\label{fme}
\ee
Thus we obtained the inverse matrix $\O^{-1}$:
\[
\Ot\O=\hat{f},~~~~~ \O^{-1}=\hat{f}^{-1}\Ot. 
\]
Before proceeding with evaluation of the matrix elements of the other operators 
in the new  basis, let us mention some usefull properties of the operator $\O$, 
and prove that that, in fact, it is the factorizing operator in a sense of 
the defenition (\ref{fff}): $\O=F^{-1}$.    
First, $\O$ and $\O^{-1}$ are the triangular matrices (upper triangular as $A(t)$). 
Second, the pseudovacuum state is an eigenstate of $\O$ ($\O^{-1}$) with the 
eigenvalue equal to unity. In general we have the following equations for 
arbitrary number of particles $n$:
\[
\O|\{00..0\}_{N-n}\{11..1\}_n\ra =1~|\{00..0\}_{N-n}\{11..1\}_n\ra
\]
\[
\la\{11..1\}_{n}\{00..0\}_{N-n}|\Ot=1~\la\{11..1\}_{n}\{00..0\}_{N-n}|,   
\]
and the same formulas for the inverse operators. 
From this formulas one can already suspect that $\O$ is the factorizing operator. 
Indeed for the particular permutation (\ref{tn}) $\sigma(\n)$
the factorizing condition is represented as 
$\O(\O^{\sigma(\n)})^{-1}=\O(\n)$, where $\O(\n)=T_{n_1}..T_{n_M}$ and is fulfilled 
at least for the state $|\n\ra$ since due to the last formulas 
$\O^{\sigma(\n)})^{-1}|\n\ra=|\n\ra$. The rigorous proof goes as follows. We  construct
the operator that acting on the state $|\n\ra$ produces the state $\O(\n)|\n\ra$. 
It is easy to see that the operator that fulfills the above requirement is: 
\be
\O = \F_1 \F_2 \ldots \F_N, ~~~~~\F_i = (1-\hat{n}_i)+T_i \hat{n}_i,  
\label{f}
\ee
where $\hat{n}_i$ is the operator of the number of particles (spin up) at the 
given site $i$. The operators $\F_i$ entering (\ref{f}) do not commute 
and their ordering in eq.(\ref{f}) is important. To prove the factorizing property 
of this operator it is sufficient to consider only one particular permutation, 
say the the permutation $(i,i+1)$ since all the others can be obtained as a 
superposition of these ones for different $i$. We will  show in the Appendix A that 
\be
\O=S_{i+1,i}\O^{(i,i+1)}, 
\label{ii}
\ee
Evidently, in contrast to (\ref{fff}), for any 
transmutation $\sigma\in S_N$  we will obtain only one operator 
$R^{\s}_{1...N}$ on the left of the operator $\O$. Thus it is proved that $\O$ is 
the factorizing operator in a sense of eq.(\ref{fff}). 

\vspace{0.2in}

         {\bf 3. Construction of the matrix elements.}

\vspace{0.2in}

   In this section we calculate the matrix elements of $B(t)$ and $C(t)$ - operators 
in the F - basis: $B^{F}(t)=FB(t)F^{-1}=\O^{-1}B(t)\O$ (and the same for $C(t)$). 
The general scheme to perform the calculations is to use the formalism developed in 
the previous section, which leads to the following chain of equations: 
\[
B(t)|\phi(\n)\ra = B(t)\O|\n\ra = \O B^F(t)|\n\ra = 
\sum_{\x}\phi(\x,t,\n)\O|\n,\x\ra. 
\]
where $|\n,\x\ra$ is the state corresponding to the 
new set of coordinates with an extra spin-up 
at the site $\x$. Thus acting by the operator $B(t)$ on an eigenstate (\ref{bb}) 
\be
B(t) B(\xi_{n_1})...B(\xi_{n_M}) \vac = \sum_{\x} \left( 
B(\xi_{\x})B(\xi_{n_1})...B(\xi_{n_M}) \vac \right) \phi(\x,t,\n), 
\label{def}
\ee
we see that $\phi(\x,t,\n)$- is exactly the matrix element in the new basis. 
To get the single term in the sum in eq.(\ref{def}), we act by the operator 
$A(\xi_{\x})$ ($\x\neq n_k$) at both sides of this equation. Using again the 
property $A(\xi_i)|0\ra=0$ and eq.(\ref{ab}) we get for the left-hand side 
of (\ref{def}) 
\[
A(\xi_{\x})B(t)B(\xi_{n_1})...B(\xi_{n_M})|0\ra = 
 -\left(\b(t-\xi_{\x})/\c(t-\xi_{\x})\right) 
B(\xi_{\x})A(t)B(\xi_{n_1})...B(\xi_{n_M})|0\ra, 
\]
while for the right-hand side we get the single term with $B(\xi_{\x})$, 
which can be easily evaluated using again eq.(\ref{ab}) and the formula 
\[
A(\xi_{\x})B(\xi_{\x})|0\ra = \left( \prod_{\a\neq\x}
\c(\xi_{\a}-\xi_{\x}) \right) B(\xi_{\x})|0\ra , 
\]
which can be proved by direct computations. After the cancellation of similar 
terms at both sides of eq.(\ref{def}) we get 
\[
\phi(\x,t,\n)=-\left(\b(t-\xi_{\x})/\c(t-\xi_{\x})\right) 
\prod_{\a\neq n_k} \c(\xi_{\a}-t) 
\left( \prod_{\a\neq n_k} \c(\xi_{\a}-\xi_{\x}) \right)^{-1}. 
\]
Then, using the equality 
\[
-\frac{\b(t-\xi_{\x})}{\c(t-\xi_{\x})}\c(\xi_{\x}-t)= \b(\xi_{\x}-t),
\]
we finally obtain in the operator form: 
\be
B^{F}(t)=\sum_{\x}\sigma_{\x}^{\dagger}~\b(\xi_{\x}-t)\prod_{\a\neq\x} 
\left( \begin{array}{ll}
\c(\xi_{\a}-t)(\c(\xi_{\a}-\xi_{\x}))^{-1},     & \a \neq n_k \\
    1,                                          & \a = n_k 
\end{array} \right).  
\label{bf}
\ee
With this expression one can prove the equation 
$\prod_i B^{F}(\xi_{n_i})\vac=|\n\ra$ which is consistent with  
the formulas of the previous section. 

   For the operator $C(t)$ proceeding in a similar way and using the 
relation (\ref{bc}), we get 
\[
C(t) B(\xi_{n_1})...B(\xi_{n_M})|0\ra = 
B(\xi_{n_1})C(t)B(\xi_{n_2})...B(\xi_{n_M})|0\ra 
\]
\be 
-\left(\b(t-\xi_{n_1})/\c(t-\xi_{n_1})\right) 
D(\xi_{n_1})A(t)B(\xi_{n_2})...B(\xi_{n_M})|0\ra. 
\label{d}
\ee
Using the symmetry between $n_1,...n_M$, we can concentrate on the 
term which describes the flipping of the spin on the $n_1$- site and 
consider only the second term in the right-hand side of the equation 
(\ref{d}). Commuting $A$ and $D$ operators to the right in (\ref{d}) and 
denoting $\xi_{n_1}=\xi_{\x}$ we obtain similarly to the previous case 
\be
C^{F}(t)=\sum_{\x}\sigma_{\x}^{-}~\b(\xi_{\x}-t)\prod_{\a\neq\x} 
\left( \begin{array}{ll}
\c(\xi_{\a}-t),                   & \a \neq n_k \\
(\c(\xi_{\x}-\xi_{\a}))^{-1},     & \a = n_k 
\end{array} \right).  
\label{cf}
\ee
The operators (\ref{bf}) and (\ref{cf}) are quasilocal i.e. they describe the 
flipping of the spin on a single site with the amplitude depending on the 
positions of spins on all the other sites of the chain. The operator 
$D^F(t)$ can be found using either the same method or the quantum determinant 
relation and has a (quasi)bilocal form. 

\vspace{0.2in}

           {\bf 4. Calculation of the scalar products.}

\vspace{0.2in}

  In this section we use the developed formalism to obtain the expressions for  
the correlation functions for the spin chains. We use the factorizing operator 
and, in particular, the expression for its matrix elements (\ref{me}) to obtain 
the expression for the general correlation function (scalar product) 
\[
S_M(\{\l\},\{t\})=\la 0|C(\l_1)C(\l_2)...C(\l_M)
                    B(t_1)B(t_2)...B(t_M)\vac ,  
\] 
where $\{\l\}$ and $\{t\}$ are two arbitrary sets of parameters (not necessarily 
satisfying the Bethe Ansatz equations). The correlation function can be 
represented in the following form 
\[
\la 0|Tr_{0_1,...0_{2M}}\left(\sigma_{0_1}^{+}...\sigma_{0_M}^{+} 
\sigma_{0_{M+1}}^{-}...\sigma_{0_{2M}}^{-} 
T_{0_1}\ldots T_{0_{2M}} \right)  \vac 
\]
The auxiliary spaces $0_1,\ldots 0_{2M}$ can be considered as a lattice 
consisting of $2M$ sites with the corresponding spectral parameters 
$\l_1,\ldots\l_M,t_1,,\ldots t_M$. Rearranging the basic $S$- matrices 
entering the product of the monodromy matrices we arrive at the operator 
\[
\tilde{T}_1 \ldots \tilde{T}_N,
\]
where the new monodromy matrices act in the auxialiary space instead of the 
original quantum space:
\[
\tilde{T}_n = S_{n0_1}S_{n0_2} \ldots S_{n0_{2M}}.
\]
Obviously using this matrices the correlator can be 
represented as the following matrix element 
in the new quantum space $0_1,...0_{2M}$:  
\[
\la \{00..0\}_M\{11..1\}_M|\tilde{A}(\xi_1)\ldots\tilde{A}(\xi_N) 
|\{11..1\}_M\{00..0\}_M \ra  
\]
(we use the symmetry $A(t)\leftrightarrow D(t)$, $0\leftrightarrow 1$ here). 
Transforming the operators $\tilde{A}(\xi_i)$ to the F - basis we find 
at $\xi_i=0$: 
\[ 
S_M = \sum_{\n} \la \{00..0\}_M\{11..1\}_M|\O|\n\ra
\la\n| \left(\tilde{A}^{F}(0))\right)^N |\n\ra
\la\n| \O^{-1} |\{11..1\}_M\{00..0\}_M \ra . 
\]
The sum is over the states labeled by the positions of M particles  
$\n=n_1\ldots n_M$ on a lattice consisting of $2M$ sites with the 
inhomogeneity parameters $\l_1\ldots\l_M,t_1\ldots t_M$. We use  
$\O^{-1}=\hat{f}^{-1}\Ot$ and the representations (\ref{me}), (\ref{mec}) 
for the matrix elements in the last formula. Let us denote by 
$\mu_1,\ldots\mu_M$ the parameters (from the set $(\{\l\},\{t\})$) 
corresponding to the sites $n_1\ldots n_M$, and by $\nu_1,\ldots\nu_M$ 
the rest of the parameters so that $\{\l\}\cup\{t\}=\{\mu\}\cup\{\nu\}$. 
The first matrix element in the formula for
$S_M$ is equal to 
\be
\la \{00..0\}_M\{11..1\}_M| B(\mu_1)\ldots B(\mu_M) \vac =
\la\{11..1\}_M| B^{\prime}(\mu_1)\ldots B^{\prime}(\mu_M) \vac, 
\label{z}
\ee
where $B^{\prime}(\mu)$ are the same operators difined on the lattice 
consisting of $M$ sites with the inhomogeneity parameters $t_1\ldots t_M$. 
The second matrix element can be reduced to the same expression with the 
parameters $\l_1\ldots \l_M$ (using the symmetry $C\leftrightarrow B$, 
$0\leftrightarrow 1$). Then using the formula for the matrix elements of the 
operator $\hat{f}^{-1}$ (see eq.(\ref{fme})) we finally obtain the 
expression: 
\be
S_M(\{\l\},\{t\})=\sum_{n_1,..n_M}\left(\prod_j a(\nu_j)\right)  
\Phi_{M}(t,\mu)\Phi_{M}(\l,\mu)
\prod_{i,j} \frac{1}{\c(\mu_i-\nu_j)}, 
\label{final}
\ee
where we denoted by $\Phi_{M}(\xi,t)$ the functions in the right hand side of 
eq.(\ref{z}). The determinant representation of this function is well known 
(see for example \cite{K1}, \cite{Coker}): 
\be
\Phi_{M}(\xi,t)=\frac{\prod_{i,j}(t_i-\xi_j)}{\prod_{i<j}(t_i-t_j)
\prod_{j<i}(\xi_i-\xi_j)}\det_{ij}\left(\frac{\eta}{(t_i-\xi_j)
(t_i-\xi_j+\eta)}\right)
\label{satur}
\ee
for the rational case. Here $\{\xi\}$ are the inhomogeneity parameters 
and $\{t\}$ are the arguments of $B$- operators (see eq.(\ref{z})). 
We present the basic  properties  of this function 
used in the calculations in the Appendix B.  
The functions $a(\nu)$ in eq.(\ref{final}) are exactly the functions 
defined above $a(\nu)=\prod_{\a}\c(\xi_{\a}-\nu)$ while in the rest 
of this formula due to the definition of the matrices $\tilde{T}_i$ 
one should interchange the arguments in the functions $\c^{-1}(\nu_i-\mu_j)$ 
which is taken into account in eq.(\ref{final}) 
(or one could make the replacement $\eta\rightarrow-\eta$).  
Using the properties of this functions one can represent the general formula 
(\ref{final}) in a different way:  
\[
\sum_{m=0}^{M}\sum_{k,n}\left(\prod \left( a(\l_n)a(t_k)\right)\right)
\Phi_{m}(t_k,\l_{\beta})\Phi_{M-m}(\l_{n},t_{\a})
\prod \frac{1}{\c(\l_{\beta}-\l_n)} \frac{1}{\c(t_{\a}-t_k)}
\frac{1}{\c(t_{\a}-\l_n)} \frac{1}{\c(\l_{\beta}-t_k)},  
\]
where the sum is over the two sets $k_1,...k_m$ and $n_1,...n_{M-m}$. 
We used the following simplified notations in this formula. We devided 
the set $\{t\}$ into two subsets $\{t\}=\{t_k\}\cup\{t_{\a}\}$ where 
$\{t_k\}=(t_{k_1}...t_{k_m})\in\{\nu\}$,  
$\{t_{\a}\}\in\{\mu\}$  and analogously $\{\l\}=\{\l_n\}\cup\{\l_{\beta}\}$, 
$\{\l_n\}=(\l_{n_1}...\l_{n_{M-m}})\in\{\nu\}$, $\{\l_{\beta}\}\in\{\mu\}$.  
The products in the last formula are over the indices labeling the elements 
of the corresponing sets. One can further rewrite this formula using 
the expressions for the functions $\Phi_m$ to get the formula which leads 
to the determinant representation for $S_M$ after the special dual fields 
are introduced \cite{IK}.

Let us derive the scalar product of the Bethe eigenstate 
with an arbitrary dual state along the lines of ref.\cite{S} starting from 
the formula (\ref{final}). The set of the parameters $\{t\}$ obey the 
Bethe equations  so for each $a(t_i)$ in the sum (\ref{final}) one can 
substitute the function 
\[
f(t_i)=\prod_{\a\neq i}\frac{\c(t_{\a}-t_i)}{\c(t_i-t_{\a})}=
\prod_{\a\neq i}\frac{\phi(t_{\a}-t_i-\eta)}{\phi(t_{\a}-t_i+\eta)}. 
\]
Then the idea is that one can calculate the sum in (\ref{final}) for an 
arbitrary smooth function $a(\l)$, which behaves at least as a constant 
at infinity, used for the terms $a(\l_j)$ 
(not necessarily equal to $a(\l)=\prod_{\a}\c(\xi_{\a}-\l)$). 
In that case the sum has the simple poles in the variables 
$\{t\}$ and $\{\l\}$ at the points $t_i=\l_j$ and don't have any other poles,  
for example the poles at $\l_i=\l_j$, which exist if, according to \cite{S} 
one considers $a(\l_j)$ as an independent variables. These poles was  
not considered in \cite{S} (actually only the behaviour of the function 
in the parameters $t_i$ was considered - obviously there is no poles at 
$t_i=t_j$ if the Bethe Ansatz equations are taken into account). 
One can use the symmetry of $S_M$ in $\{t\}$ and $\{\l\}$ and single out the 
residue at $\l_1\rightarrow t_1$ (which is contained in the term with 
$\l_1\in\{\nu\}$, $t_1\in\{\mu\}$ and vice versa):  
\[
S_M \left( \{\l\},\{t\},a(\nu_j) \right)  \vert_{\l_1\rightarrow t_1}
\rightarrow \eta \frac{a(\l_1)-f(t_1)}{t_1-\l_1} 
\prod_{\a\neq 1} \frac{1}{\c(t_{\a}-t_1)} \frac{1}{\c(\l_{\a}-t_1)}
S_{M-1} \left( \{\l\}',\{t\}',a'(\nu_j) \right), 
\]
where $\{\l\}'$, $\{t\}'$ do not contain  $\l_1$, $t_1$ and 
\[
a'(\nu_j)=a(\nu_j) \frac{\c(\nu_j-t_1)}{\c(t_1-\nu_j)}. 
\]
The functions $f(t_i)$ entering the sum $S_{M-1}$ are also modified 
so that the terms with $t_1$ are absent in their definition. 
Also the behaviour of $S_M\sim 1/\l_1$ at $\l_1$ going to infinity 
should be taken into account.   
It can be easily proved that the same recurrence relation at  
$\l_1\rightarrow t_1$ is obeyed by the following expression: 
\[
S_M(\{\l\},\{t\})=\frac{1}{\prod_{i<j}(t_i-t_j)\prod_{j<i}(\l_i-\l_j)}
\det_{ij}(M_{ij}(t,\l))
\]
\be
M_{ij}(t,\l) = \frac{\eta}{(t_i-\l_j)} 
 \left( a(\l_j)\prod_{\a\neq i}(t_{\a}-\l_j +\eta) - 
 \prod_{\a\neq i}(t_{\a}-\l_j - \eta) \right),  
\label{mbethe} 
\ee
where at the end of calculations one should take the actual 
form of the function $a(\l)$. So we proved that for an arbitrary 
function $a(\l)$ the residues of the only existing poles at 
$t_i=\l_j$ satisfy the the same recursion relation in both 
formulas. Since the function (in each variable) is completely 
determined by the positions of (all) poles and the corresponding 
residues (along with the behaviour at the infinity) the two 
functions (\ref{mbethe}) and (\ref{final}) should coincide for 
an arbitrary function $a(\l)$.   
To complete the proof, one should check the formula (\ref{mbethe}) 
for $M=1,2$. Equivalently, one can check the coefficients 
corresponding to the terms $\prod_{i=1}^{M}a(\l_i)$ and 
$\prod_{i=1}^{M}a(t_i)=1$ in eq.(\ref{mbethe}) which are 
respectively 
\[
\Phi_M(\l,t)\prod_{i,j}\frac{1}{\c(t_i-\l_j)},~~~~~~~  
\Phi_M(t,\l)\prod_{i,j}\frac{1}{\c(\l_i-t_j)} 
\]
in agreement with (\ref{final}). 
The formula (\ref{mbethe}) is written for the rational case. 
The generalization to the case of XXZ chain is obvious. 
This formula can also be represented through the Jacobian as: 
\[
S_M(\{\l\},\{t\})=(-1)^M
\frac{\prod_{i,j}(t_i-\l_j)}{\prod_{i<j}(t_i-t_j)\prod_{j<i}(\l_i-\l_j)} 
\det_{ij}\left(\frac{\partial}{\partial t_i}\Lambda(\l_j;\{t_{\a}\})\right). 
\]
Here the sign $(-1)^M$ can be absorbed into the product 
$\prod_{i\neq j}(\l_i-t_j)=(-1)^{M}\prod_{i\neq j}(t_i-\l_j)$. 
The orthogonality of two different Bethe eigenstates was shown 
in ref.(\cite{KS}). From this expression taking the limit 
$\l_i\to t_i$ one can easily obtain 
the formula for the norm of the Bethe eigenstate: 
\[
N_M(t)=\eta^M\frac{\prod_{i\neq j}(t_i-t_j+\eta)}{\prod_{i\neq j}(t_i-t_j)}
\det_{ij}\left[- \frac{\partial}{\partial t_j}
\ln \left(\frac{a(t_i)}{f(t_i)}\right) \right], 
\]
where $f(t_i)=\prod_{\a\neq i}\c(t_{\a}-t_i)/\c(t_i-t_{\a})$.  
Note that the matrix 
$N_{ij}=-\partial/\partial t_j \ln\left(a(t_i)/f(t_i)\right)$ 
can also be represented in the form: 
\[
N_{ij}=\frac{2\eta}{(t_{ij}+\eta)(t_{ij}-\eta)},~~~~i\neq j, 
\]
\[
N_{ii}= - \frac{\partial}{\partial t_i} \ln\left( a(t_i)) \right) -  
  \sum_{\a\neq i} \frac{2\eta}{(t_{\a i}+\eta)(t_{\a i}-\eta)}, 
\]
where $t_{ij}=t_i-t_j$. The first term in $N_{ii}$ should be understood as 
$-\left(\ln(a(t))\right)'(t=t_i)$. 
To calculate the correlators in the continuum limit it can be  
useful to represent the matrix $N$ as a product of two matrices as 
\[
N_{ij}=\sum_{\a}\left(\delta_{i\a}+(1-\delta_{i\a})
N_{i\a}(N_{\a\a})^{-1}\right) 
\left(\delta_{\a j}N_{jj}\right), 
\]
where $N_{ii}\sim R(t_i)$ - the density of the variables $t_i$ 
in the continuum limit. 
The determinant expressions for the formfactors can be rather 
straightforwardly obtained from the above formulas using the explicit 
solution of quantum inverse scattering problem presented in ref.\cite{KMT}. 

In conclusion, for the calculation of correlation functions using 
the formfactors, it seems necessary to develop the method to 
evaluate the dependence of the formfactors of local operators 
($\sigma_{i}^{\pm,z}$) on the quantum numbers (quasiparticle numbers) 
characterizing the ground and the excited states, the problem 
which was not solved by now. The physical results have been obtained 
only for the simplest formfactors of the operator $\sigma^z$ 
for the massles \cite{KMT} and massive \cite{IKMT} (Baxter's 
formula for spontaneous magnetization) regimes of the XXZ chain.

\vspace{0.1in}

{\bf Acknowledgements }

The author is deeply indebted to the stuff of INR Theory Division   
for the support.

\vspace{0.3in}

         {\bf Appendix A. }   

\vspace{0.1in}

We give in this appendix the proof that the operator (\ref{f}) introduced 
in the text is the factorizing operator:  
\be
\O = \F_1 \F_2 \ldots \F_N, ~~~~~\F_i = (1-\hat{n}_i)+T_i \hat{n}_i,  
\label{a1}
\ee
As it was explained in section 2 it is sufficient to prove the property   
$\O=S_{i+1,i}\O^{(i,i+1)}$, where by the definition 
$\O^{(i,i+1)}=\O_{12...i+1,i,..N}$ for the one particular permutation $(i,i+1)$.  
Consider the component $\F_i\F_{i+1}$ in eq.(\ref{a1}). Substituting in this term 
$T_i=S_{i+1,i}T_{i}^{(i+2...N)}$, where 
\[
T_{i}^{(i+2...N)}=S_{i+2,i}\ldots S_{Ni}, 
\]
we obtain: 
\[
\F_i\F_{i+1}= (1-\nn_i)(1-\nn_{i+1}) + 
S_{i+1,i}T_{i}^{(i+2...N)}T_{i+1}^{(i+2...N)}\nn_i\nn_{i+1} + 
\]
\be
S_{i+1,i}T_{i}^{(i+2...N)}\nn_i(1-\nn_{i+1}) + 
T_{i+1}^{(i+2...N)}\nn_{i+1}(1-\nn_i). 
\label{a2}
\ee
Moving $S_{i+1,i}$ to the right in the second term of eq.(\ref{a2}) using the 
Yang- Baxter equation, using the unitarity of the $S$- matrix and taking 
into account that $S_{i,i+1}\nn_i\nn_{i+1}=\nn_i\nn_{i+1}$ we rewrite (\ref{a2}) 
in the following form: 
\[
S_{i+1,i} ~ (~   (1-\nn_i)(1-\nn_{i+1}) + 
S_{i,i+1}T_{i+1}^{(i+2...N)}T_{i}^{(i+2...N)}\nn_i\nn_{i+1} + 
\]
\be
S_{i,i+1}T_{i+1}^{(i+2...N)}\nn_{i+1}(1-\nn_{i}) + 
T_{i}^{(i+2...N)}\nn_{i}(1-\nn_{i+1}) ~)~      
\label{a3}
\ee
Comparing eq.(\ref{a2}) and eq.(\ref{a3}), we see that the expression in the
brackets in (\ref{a3}) coincide with the right-hand side of (\ref{a2}) up to
the interchange of the indices $i\leftrightarrow i+1$ (together with the 
corresponding parameters $\xi_i$, $\xi_{i+1}$). In the expression (\ref{a1}) 
the indices $i$, $i+1$ are contained only on the left from the operator
$\F_i\F_{i+1}$ in the operators $T_n$ for $n<i$: 
\[
T_n=S_{n+1,n}\ldots S_{i,n}S_{i+1,n}\ldots S_{N,n}. 
\]
Thus moving the matrix $S_{i+1,i}$ in (\ref{a3}) to the left and using the 
Yang - Baxter equation, we interchange the indices $i$, $i+1$ in the whole 
expression for the operator $\O$  (\ref{a1}) and arrive exactly to the 
formula (\ref{ii}).

\vspace{0.3in}

    {\bf Appendix B. }

\vspace{0.1in}

In this appendix we derive the expression for the function 
$\Phi_M(\xi,t)$ - the partition function of a six- vertex model 
with domain wall boundary conditions, 
and point out its properties used in the text. 
For simplicity we consider the rational case.  
We define this function using the operators $B(t)$ corresponding  
to the  monodromy matrix defined according to 
$\tilde{T}_{0}(t)=S_{01}S_{02}\ldots S_{0M}$ with the inhomogeneity 
parameters $\xi_1,\ldots\xi_M$ (that corresponds to its 
definition in the text):  
\[
\Phi_M(\{\xi\},\{t\})= \langle \{111..1\}_M|B(t_1)B(t_2)\ldots 
B(t_M)\vac.
\]
It is just the same function that appears in the formula  
(\ref{final}). The function is determined entirely by the following 
properties. 

1) At $M=1$, $\Phi_1=\eta/(t-\xi+\eta)$. 

2) $\Phi_M(\{\xi\},\{t\})$ is a symmetric function 
of the variables $\{t\}$ and separately $\{\l\}$. 

3) If one chooses the initial (non -unitary) normalization of the 
$S$- matrix (and the corresponding $B$ operators), $\Phi_M$ is 
a polynomial of degree $M-1$ in each of the variables $t_i$, $\l_i$. 
The polynomial in $t_i$ is completely fixed by its values at $M$ 
different points (for example at the points $\xi_j$).  

4) At $t_1=\xi_1$ the function reduces to 
$\Phi_{M-1}(\{\xi_i\}_{i\neq 1},\{t_i\}_{i\neq 1})$ and the same 
for all the other pairs of variables.  This can be easily seen  
using the symmetry property, by placing the corresponding variable 
$\xi_i$ to the end of the chain (at M-th site). Altenatively, one 
can use the representation of $B$- operators in the F- basis 
(\ref{bf}). 

These properties determine the function $\Phi_M$ unambigously 
(see eq.(\ref{satur}) in the text). One should mention that it 
does not have poles at $t_i=t_j$ and $\xi_i=\xi_j$ since at these  
values the two lines or coloumns in the determinant coincide.

One can also determine the function $\Phi_M(\xi,t)$ directly using 
the operators in the F- basis. This was done in ref.\cite{KMT}. 
Let us present here this proof with some technical modifications 
which could be important for the other problems of this type and will be 
used in the Appendix C.  
Consider the determinant 
\[
\det_{ij}(M_{ij})=\det_{ij}\left(\frac{1}{(t_i-\l_j)(t_i-\l_j+\eta)}\right). 
\]
First, we modify the first line of the matrix $M_{1j}$ adding 
a linear combination of the other lines so that the determinant 
$\det M$ is unchanged: 
\[
M'_{1j}=M_{1j}+\sum_{x\neq 1}C_x M_{xj}
\]
where the coefficients are chosen to be 
\[
C_x=-\prod_{\beta}\frac{(t_x-\l_{\beta}+\eta)}{(t_1-\l_{\beta}+\eta)}
\prod_{\a\neq 1,x}\frac{(t_1-t_{\a})}{(t_x-t_{\a})}. 
\]
It is necessary to calculate the following sum: 
\be
\sum_{x\neq 1}C_x M_{xj}=
-\frac{\prod_{\a\neq 1}(t_1-t_{\a})}{(t_1-\l_j+\eta)}
~\sum_{x\neq 1}f(t_x)\prod_{\a\neq 1,x}\left(\frac{1}{t_x-t_{\a}}\right)
\frac{1}{(t_1-t_x)}\frac{1}{(t_x-\l_j)}, 
\label{b1}
\ee
where
\[
f(t_x)=
\prod_{\beta\neq j}\frac{(t_x-\l_{\beta}+\eta)}{(t_1-\l_{\beta}+\eta)}.
\]
To calculate the sum in eq.(\ref{b1}) consider the integral in the complex 
plane over the circle of large radius which is equal to zero due to the 
behaviour of the integrand at infinity: 
\[
\oint dz\frac{f(z)}{(z-t_1)(z-\l_j)\prod_{\a\neq 1}(z-t_{\a})}=0,
\] 
where 
\[
f(z)=
\prod_{\beta\neq j}\frac{(z-\l_{\beta}+\eta)}{(t_1-\l_{\beta}+\eta)}.
\]
The sum of the residues at $z=t_{\a}$ is equal to the sum in (\ref{b1}), 
so it is easily calculated to be 
\[
\frac{f(t_1)}{(t_1-\l_j)\prod_{\a\neq 1}(t_1-t_{\a})} + 
\frac{f(\l_j)}{(\l_j-t_1)\prod_{\a\neq 1}(\l_j-t_{\a})}. 
\]
Thus combining all terms we obtain the following expression for the 
new elements of the first line: 
\be
M'_{1j}=\frac{1}{(t_1-\l_j)(t_1-\l_j+\eta)}
\prod_{\a\neq 1}\frac{(t_1-t_{\a})}{(\l_j-t_{\a})}
\prod_{\beta\neq j}\frac{(\l_j-\l_{\beta}+\eta)}
{(t_1-\l_{\beta}+\eta)}. 
\label{b2}
\ee
The same formula could be obtained for the first column of the matrix 
$M_{ij}$. Note also that taking the coefficiens $C_x$ in a different 
way we could obtained the other expressions of this type. For example, 
taking the coefficients 
\[
M'_{i1}=M_{i1}+\sum_{x\neq 1}C_x M_{ix},~~~~~
C_x = - \prod_{\a}\frac{(\l_x-t_{\a})}{(\l_1-t_{\a})} 
\prod_{\beta\neq 1,x}\frac{(\l_1-\l_{\beta})}{(\l_x-\l_{\beta})},
\]
we obtain the new expression of the first column of the matrix $M'$ 
in the form: 
\[
M'_{i1}= 
\frac{1}{(t_i-\l_1)(t_i-\l_1+\eta)}
\prod_{\a\neq i}\frac{(t_i-t_{\a}+\eta)}{(\l_1-t_{\a})}
\prod_{\beta\neq 1}\frac{(\l_1-\l_{\beta})}
{(t_i-\l_{\beta}+\eta)},
\]
which is different from (\ref{b2}). Note also an interesting property 
of the function $\Phi_M(\xi,t)$, which can be seen from these 
formulas. Namely the function $\Phi_M$ is equal to zero at the points 
$\xi_i=\xi_j\pm\eta$. We will use this property in the Appendix C.  

Now the derivation of the formula (\ref{satur}) is straightforward. 
Using the operators $B(t)$ in the F- basis and acting on the 
vacuum state by one of the commuting operators, say the operator 
$B^{F}(t_1)$ first, we get the formula 
\[
\Phi_M(\xi,t)=\sum_{i=1}^{M}f_{\xi_i}(t_1)F(\xi,t),~~~~~~
f_{\xi_i}(t_1)=\b(t_1-\xi_i)\prod_{\a\neq i}
\frac{\c(t_1-\xi_{\a})}{\c(\xi_i-\xi_{\a})}
\]
(we change the arguments in comparison with eq.(\ref{bf}) due to 
our definition of the function $\Phi_M$), where the function 
$F(\xi,t)$ corresponds to the action of the other operators $B^F(t_j)$. 
Each term in this sum corresponds to the occupied site $i$, so 
due to the form of the operators $B^F$ (\ref{bf}) the action of the other 
operators 
\[
B^F(t_M)\ldots B^F(t_3)B^F(t_2)
\]
on this state does not contain the variable $\xi_i$. 
Thus we single out the dependence of the function on the variables 
$\xi_i$ and $t_1$, and the function $F(\xi,t)$ is equal to
\[
F(\xi,t)= \Phi_{M-1}(\{\xi_{\beta}\}_{\beta\neq i},
 \{t_{\a}\}_{\a\neq 1}),
\]
and does not contain these variables. 
Apart from the factor in front of the determinant that procedure  
corresponds to the first line development of the determinant in 
(\ref{satur}). One can check that starting from the last formula  
we get exactly the first line development of the determinant 
with the matrix elements $M'_{1i}$ instead of $M_{1i}$. 
In fact, assuming that the function $\Phi_{M-1}$ is given by the 
formula (\ref{satur}) we obtain from the last formula: 
\[
\Phi_{M}(\xi,t)=\frac{\prod_{i,j}(t_i-\xi_j)}{\prod_{i<j}(t_i-t_j)
\prod_{j<i}(\xi_i-\xi_j)} \sum_i (-1)^{i} 
\det(M_{\a\beta})\vert_{\a\neq 1,\beta\neq i}^{(M-1)}(\xi,t)  
\]
\[
\left\{ \frac{\eta}{(t_1-\xi_i)(t_1-\xi_i+\eta)}
\prod_{\a\neq 1}\frac{(t_1-t_{\a})}{(\xi_i-t_{\a})}
\prod_{\beta\neq i}\frac{(\xi_i-\xi_{\beta}+\eta)}
{(t_1-\xi_{\beta}+\eta)} \right\}.
\]
Comparing this formula with (\ref{b2}) we see that the last term 
in the brackets is exactly equal to $M'_{1i}$ and we obtain the 
exactly the determinant of the matrix $M'$.  
Since we have shown that the two determinants are equal   
this procedure gives us the recurrence relation which proves the 
formula (\ref{satur}).

\vspace{0.3in}
              {\bf Appendix C.}
\vspace{0.1in}

  In this appendix we will establish the direct correspondence 
between the calculations of the correlators in the F- basis 
and the previous calculations \cite{K1} based on the commutation 
relations (or, alternatively using the method of the present paper). 
Namely, starting from the expression of the scalar product in the 
F- basis we obtain the general formula (\ref{final}) for an 
arbitrary parameters $\{t\}$, $\{\l\}$. 
  To obtain the expression of the scalar product in the F- basis 
we insert the complete set of states labeled by the coordinates
$\{x\}=(x_1,\ldots x_M)$ as follows: 
\[
S_M=\sum_{\{x\}}~\la0|C^F(\l_1)\ldots C^F(\l_M)|\{x\}\ra  
  \la\{x\}|B^F(t_1)\ldots B^F(t_M)|0\ra.
\]
One can use the expressions for $B^F$, $C^F$ from Section 3 to find 
the expression for $S_M$ (which is possible due to the quasilocal 
structure of these operators).  
However the simplest way is to reduce the 
matrix elements to the scalar products of the usual form: 
\[
S_M(\{\l\},\{t\})=\sum_{x_1,..x_M}
\left(  \prod_{i,j,~\a_i\neq x_j} 
\frac{1}{\c(\xi_{\a_{i}}-\xi_{x_j})} \right)
\la0|C(\l_1)\ldots C(\l_M)B(\xi_{x_1})\ldots B(\xi_{x_M})|0\ra 
\]
\[
\hspace{2.5cm}
\la 0|C(\xi_{x_1})\ldots C(\xi_{x_M}) B(t_1)\ldots B(t_M)|0\ra.
\]
This formula is obtained by 
transforming the operators $B$ and $C$ back to the initial basis 
and using the formulas of Section 2. The last scalar products can 
be easily evaluated using eq.(\ref{final}). 
Proceeding in this way we obtain the following expression:
\[
S_M(\{\l\},\{t\})=\prod_{i=1}^M a(t_i)\prod_{j=1}^M a(\l_j)
\hspace{2.5cm}
\]
\be
\sum_{\{x\}}~\prod\frac{1}{\c(\xi_{\a}-\xi_x)}
\left( \Phi_M(\l,\xi_x) \prod\frac{1}{\c(\xi_x-\l)} \right)
\left( \Phi_M(t,\xi_x) \prod\frac{1}{\c(\xi_x-t)} \right), 
\label{fbas}
\ee
where we have introduced the sets of parameters 
$\{\xi\}=\{\xi_x\}\cup\{\xi_{\a}\}$ with  
$\{\xi_x\}=(\xi_{x_1}\ldots\xi_{x_M})$ and the products are 
over all elements of the corresponding sets 
of parameters ($\{\xi_x\}$, $\{\xi_{\a}\}$, $\{t\}$, $\{\l\}$).   
The function given by the product of two brackets in 
eq.(\ref{fbas}) has only simple poles at $\xi_{x_i}=t_j$ 
and $\xi_{x_i}=\l_j$. We remind that the function 
$\Phi_M(\l,\xi_x)\prod(\xi_x-\l+\eta)$ has no poles in $\xi_{x_i}$. 
The behaviour of this function at the infinity is 
$\sim1/\xi_{x_i}^2$ for each $\xi_{x_i}$. 
To take the sum we apply the same method as in the Appendix B. 
The problem is that now we have to sum over $M$ coordinates 
$x_1,\ldots x_M$ with the condition $x_i\neq x_j$. 
First, as an example, let us consider the following simplified 
sum which is quiet similar to that in eq.(\ref{fbas}): 
\[
S = \sum_{x_1,..x_M}\left(\prod_{i,j,~\a_i\neq x_j} 
\frac{1}{(\xi_{\a_{i}}-\xi_{x_j})}  
\prod_{i,j}\frac{1}{(\xi_{x_i}-\l_j)}\right). 
\]
To calculate this sum let us multiply both the denominator 
and the numerator by 
$\prod_{i\neq j}(\xi_{x_i}-\xi_{x_j})$. Then we obtain 
the following sum: 
\[
\sum_{x_1,..x_M}~ \frac{\prod_{i\neq j}(\xi_{x_i}-\xi_{x_j})}
{\prod_{\a\neq x_1}(\xi_{\a}-\xi_{x_1})\ldots 
 \prod_{\a\neq x_M}(\xi_{\a}-\xi_{x_M})} 
\prod_{i,j}\frac{1}{(\xi_{x_i}-\l_j)}, 
\]
where the sum is still over the configurations with $x_i\neq x_j$. 
However one can relax this condition since in the denominator 
we have the function which is not equal to zero at $x_i=x_j$, 
while the numerator is equal to zero at $x_i=x_j$. Then the sum 
is represented by the product of the sums of the type (\ref{b1}) 
(with some degree of $\xi_{x_i}$ in the numerator) which can be 
easily calculated by the same method. 
In fact, by means of the polynomial 
\[
\prod_{i\neq j}(\xi_i-\xi_j)=\sum_{k_1,\ldots k_M}
C_{k_1,\ldots k_M}\xi_{1}^{k_1}\ldots \xi_{M}^{k_M}, 
\]
($k_i\in Z^{+}$) the sum is represented as 
\[
S= \sum_{k_1,\ldots k_M}C_{k_1,\ldots k_M}
\left(\sum_{x_1}\frac{\xi_{x_1}^{k_1}}{f(\xi_{x_1})}\right)
\ldots 
\left(\sum_{x_M}\frac{\xi_{x_M}^{k_M}}{f(\xi_{x_M})}\right), 
\]
where
\[
f(\xi_{x_i})= \prod_{\a\neq x_i}(\xi_{\a}-\xi_{x_i})
\prod_{j=1}^M (\l_j-\xi_{x_i}). 
\]
Each of the sums over $x_i$ can be 
rewritten as in the Appendix B:
\[
\sum_{x_1}\frac{\xi_{x_1}^{k_1}}{f(\xi_{x_1})}
=\sum_{x_1}\frac{\l_{x_1}^{k_1}}{\tilde{f}(\l_{x_1})},
~~~~~\tilde{f}(\l_{x_1})=
\prod_{\a}(\xi_{\a}-\l_{x_1})
\prod_{j\neq x_1}(\l_j-\l_{x_1}). 
\] 
Then we obtain the product of the other sums 
of this type and the procedure can be 
repeated in the opposite direction. 
Thus we obtain the following 
simple result for this sum: 
$S=\prod_{\a=1}^N\prod_{i=1}^M(\xi_{\a}-\l_i)^{-1}$. 
Now using the same method one can calculate directly the sum 
(\ref{fbas}). Schematically, for each of the variables 
$\xi_{x_i}$ one can take either the pole at $t_i$ or the pole at 
$\l_j$. Taking the pole at $t_i$ we get $1/a(t_i)$ which cancels 
the corresponding factor in front of the sum in (\ref{fbas}). 
Using the notations of the formula (\ref{final}), that means that 
the corresponding $t_i$ belongs to the set $\{\mu\}$. We should also 
obtain the factor $\c^{-1}(\mu-\nu)$ arriving to the formula 
(\ref{final}). In fact, using the same method as before we 
represent this sum as: 
\[
\sum_{x_1\ldots x_M;~x_i\neq x_j}~ \prod_{j=1}^M
\left(\frac{1}{\prod_{\a\neq x_j}\c(\xi_{\a}-\xi_{x_j})
\prod_{i}\c(\xi_{x_j}-t_i)\prod_{i}\c(\xi_{x_j}-\l_i)}\right) 
\]
\[
\prod_{i\neq j}(\xi_{x_i}-\xi_{x_j})
\prod_{i\neq j}(\xi_{x_i}-\xi_{x_j}+\eta)^{-1}
\Phi_M(t,\xi_x)\Phi_M(\l,\xi_x)
\]
The function in the second line of this formula $F(\xi_x,t,\l)$ -  
is the function of $3M$ variables. The function $F$ has the 
following properties. 1) $F$ is equal to zero at $x_i=x_j$, so we 
can relax the condition $x_i\neq x_j$ and sum over all $x_i$ 
independently. 2) The function $F(\xi_x,t,\l)$ does not have 
poles at the points $\xi_{x_i}=\xi_{x_j}\pm\eta$ since the function 
$\Phi_M(t,\xi_x)$ is equal to zero at these points. This can be seen 
from the representation given by the last equation of the Appendix B. 
The only poles contained in the function $F$ are at 
$\xi_{x_i}=t_j-\eta$ and $\xi_{x_i}=\l_j-\eta$. 
3) The behaviour of $F$ at $|\xi_{x_i}|\to\infty$ is 
$\sim 1/\xi_{x_i}^2$. 
We can take the sum consequently over $x_1,\ldots x_M$. Taking
the sum over $x_1$ we obtain residues at some $t_i$ or $\l_i$ 
(which corresponds to the set $\{\mu\}$). One can see that 
the residues corresponding to the poles contained in the function 
$F$ are equal to zero due to the terms of the first line of the 
last formula. Then we repeat this procedure for $x_2,\ldots x_M$. 
Note that the pole we take for $x_2$ should not coincide with the 
pole we took at the previous stage (for $x_1$).  
One could formulate the same  procedure in another, equivalent way. 
Consider the function 
\[
F'(\xi_x,t,\l)=F(\xi_x,t,\l)\prod(\xi_x-t+\eta)\prod(\xi_x-\l+\eta)= 
\]
\[
\prod_{i\neq j}\c(\xi_{x_i}-\xi_{x_j})
\Phi_M(t,\xi_x)\Phi_M(\l,\xi_x)\prod_{i,j}(\xi_{x_i}-t_j+\eta)
\prod_{i,j}(\xi_{x_i}-\l_j+\eta)
\]
This function does not have poles at all and therefore is a 
polynomial. Then the sum is represented as the product of independent 
sums (as for our simplified example). 
Thus, we finally obtain the result: 
\[
\sum_{n_1\ldots n_M}\prod_{i=1}^{M}\frac{1}{a(\mu_i)}
\prod_{i,j}\frac{1}{\c(\mu_i-\nu_j)} \Phi_M(t,\mu)\Phi_M(\l,\mu),
\]
where the notations are the same as in the equation (\ref{final}). 
Thus the equation (\ref{final}) is reproduced. 
Note that one could start from the equation (\ref{final}) 
and derive the equation (\ref{fbas}) using the same method 
substituting 
$\prod_{i}a(\nu_i)=\prod_{i}a(t_i)\prod_{i}a(\l_i)/\prod_{i}a(\mu_i)$
and taking the sum over $\mu_1,\ldots\mu_M$ with 
$a(\mu)=\prod_{\a=1}^{N}\c(\xi_{\a}-\mu)$ 
(taking the residues corresponding to the poles at the points 
$\mu_{i}=\xi_{x_i}$).

  In conclusion, let us present two equivalent ways to rewrite 
the expression (\ref{final}) for the case when 
the parameters $\{t\}$ satisfy the Bethe Ansatz equations.  
Expressing the functions $a(t_k)$ and using 
the explicit form of the function $\Phi_M$ we get: 
\[
S_M(\l,t)=\frac{1}{\prod_{i<j}(t_i-t_j)}\frac{1}{\prod_{j<i}(\l_i-\l_j)}       
\sum_{k,n} (-1)^{P_k}(-1)^{P_n}  \prod(a(\l_n)) 
\]
\[
\det(t_k,\l_{\beta}) \det(\l_n,t_{\a})
\prod(\l_{\beta}-\l_n+\eta)(t_k-t_{\a}+\eta)(t_{\a}-\l_n+\eta)
(\l_{\beta}-t_k+\eta), 
\]
where again the products is over all elements of the corresponding 
sets of parameters. Here the sign factors $P_k$, $P_n$ depend on the 
sets of the coordinates $k_1,\ldots k_m$, $n_1,\ldots n_{M-m}$ in 
order to obtain the factors in front of the sum from the terms 
$\prod(t_k-t_{\a})$ and $\prod(\l_{\beta}-\l_n)$. 
We also used the simplified notations for the determinants (\ref{satur}).  
One can further rewrite this formula as follows: 
\[
S_M(\l,t)=\frac{1}{\prod_{i<j}(t_i-t_j)}\frac{1}{\prod_{j<i}(\l_i-\l_j)}       
\sum_{k,n} (-1)^{P_k}(-1)^{P_n} (-1)^{Mm} \prod(a(\l_n)) 
\]
\[
\prod (t-\l_n+\eta) \prod (t-\l_{\beta}-\eta)
\det(t_k,\l_{\beta}) \det(\l_n,t_{\a})
\prod \left( \frac{t_k-t_{\a}+\eta}{t_k-\l_n+\eta}  \right)
\prod \left(\frac{\l_{\beta}-\l_n+\eta}{\l_{\beta}-t_{\a}+\eta} \right),   
\]
where we denote for example 
$\prod(t-\l_n+\eta)=\prod_j\prod_{i=1}^{M}(t_i-\l_{n_{j}}+\eta)$. 
The other notations are the same as before.  
In this formula the poles at $t_i=\l_j$ are contained in the determinants. 
If the two last products are dropped out this formula becomes the result of 
the decomposition of the determinant in eq.(\ref{mbethe}). Then one can 
see explicitly that the residues in the poles in the formulas (\ref{final}) 
and (\ref{mbethe}) satisfy the same recurrence relations.

\end{document}